\documentclass[onecolumn,showpacs]{revtex4}

\usepackage{graphicx}
\usepackage{dcolumn}
\usepackage{bm}

\input epsf

\begin{document}

\title{\Large  THE GENERALIZED SECOND LAW OF THERMODYNAMICS OF THE UNIVERSE BOUNDED
BY THE EVENT HORIZON AND MODIFIED GRAVITY THEORIES}

\author{\bf~Nairwita~Mazumder\footnote{nairwita15@gmail.com}, Subenoy~Chakraborty\footnote{schakraborty@math.jdvu.ac.in}.}

\affiliation{$^1$Department of Mathematics,~Jadavpur
University,~Kolkata-32, India.}

\date{\today}

\begin{abstract}
In this paper, we investigate the validity of the generalized
second law of thermodynamics of the universe bounded by the event
horizon. Here we consider homogeneous and isotropic model of the
universe filled with perfect fluid in one case and in another case
holographic model of the universe has been considered. In the
third case the matter in the universe is taken in the form of
non-interacting two fluid system as holographic dark energy and
dust. Here we study the above cases in the Modified gravity, f(R)
gravity.
\end{abstract}


\maketitle

\section{\normalsize\bf{Introduction}}

After the discovery of black hole thermodynamics physicists start
speculating about the inherent link between the black hole
thermodynamics and Einstein's field equations. Jacobson first
showed the link between general relativity and thermodynamics
$[1]$. Einstein's equation were shown to be the consequences of
the proportionality of entropy and horizon area which was first to
outline the laws of thermodynamics in the field of a black hole
by Bekenstein $[2]$. Later Padmanavan $[3]$ derived the first law
of thermodynamics on the apparent horizon starting from Einstein's
equations on it. If we assume the universe as a thermodynamical
system and consider that at the apparent horizon $R=R_{A}$, the
Hawking temperature $T_{A}=\frac{1}{2 \pi R_{A}}$ and the entropy
$S_{A}=\frac{ \pi {R_{A}}^2}{G}~$, then it can be shown that the
first law of thermodynamics and the Friedmann equations are
equivalent $[4]$.\\

At present there are various observational data which strongly
suggest that the current expansion of the universe is
accelerating $[5,6]$. Now the reason for this accelerating
expansion can be described in two ways $[7,8,9,10,11,12]$. One is
to introducing the dark energy having negative pressure in the
frame work of general relativity and other is to study a modified
gravitational theory, such as f(R) gravity $[12,13,14]$ where the
action term is described by an arbitrary function f(R) of the
scalar curvature $R$. In this modified theory of gravity, instead
of Friedmann equations we have the modified Friedmann equations
$[15,16]$ which includes the powers of Ricci Scalar and its time
derivatives. In f(R) gravity the expression of the entropy is
also different from that of Einstein gravity namely, $S=f'(R)\frac{A}{4G}$ $[15]$.\\

Similar to this situation the entropy area relation may have a
logarithmic correction term i.e. $S=\frac{A}{4G}+ \alpha \ln
\frac{A}{4G}~$, where $\alpha$ is a dimension less constant. Such
a correction term is expected to be a generic one in any theory of
quantum gravity. Due to this quantum correction, the Friedmann
equations also get modified and are familiar in the form to loop
quantum gravity[17-25]. In this modified gravity theory the big
bang singularity can be described by a quantum bounce[26].\\

In the usual standard big bang model a cosmological event horizon
does not exist. But for the accelerating universe dominated by
dark energy, with equation of state $\omega_{D}\neq -1$, the
cosmological event horizon separates from that of the apparent
horizon. Wang et.al. $[27]$ have shown that by applying the usual
definition of temperature and entropy as in apparent horizon to
the cosmological event horizon, both first and second law of
thermodynamics breaks down. They have argued that the first law
is applicable to nearby states of local thermodynamic equilibrium
but event horizon reflects the global features of space-time. The
definition of thermodynamical quantities on the cosmological
event horizon in the non-static universe may not be as simple as
in the static space-time. Further, it is speculated that the
region bounded by the apparent horizon may be considered as the
Bekenstein system i.e. Bekenstein's entropy-mass bound ($S\leq 2
\pi R_{E}$) and entropy-area bound ($S \leq \frac{A}4$) are
obeyed in this region. Moreover, Bekenstein bounds are universal
and all gravitationally stable special regions with weak self
gravity should satisfy Bekenstein bounds and the corresponding
thermodynamical system is termed as Bekenstein system. But event
horizon is larger than the apparent horizon so the universe
bounded by the event
horizon is not a Bekenstein system.\\

In this paper we try to find the necessary conditions for the
validity of the generalized second law of thermodynamics on the
event horizon both modified gravity with logarithmic correction in
entropy-area relation(MGLC) and in f(R) gravity. Here we have not
assumed any explicit expression for entropy and temperature in the
event horizon but have only assumed the validity of the first law
of thermodynamics (which can be viewed as an energy
conservation relation) on the event horizon.\\

The paper is organized as follows: in section two we have taken
the homogeneous and isotropic model of the universe filled with
perfect fluid both in MGLC as well as in f(R) gravity. In the
third section we have considered the universe filled only with
holographic dark energy. The holographic dark energy model
$[27-39]$ has been constructed in the light of holographic
principle of quantum gravity theory. Then in section four, matter
in the non-interacting combination of holographic dark energy and
dust has been chosen for the universe. The conditions for the
validity of the generalized second law of thermodynamics on the
event horizon are determined both in MGLC and f(R) gravity.
Finally, the paper ends with concluding remarks in section $V$.\\

\section{\normalsize\bf{Generalized Second Law of thermodynamics of the universe filled with perfect fluid:}}

\subsection{\normalsize\bf{\underline{Modified Gravity with Logarithmic Correction in Entropy-Area Relation(MGLC)}}}

In FRW model of the universe filled with perfect fluid with
non-zero spatial curvature,  the modified Friedmann equations in
MGLC are given by [40]

\begin{equation}
\left[1+\frac{ \alpha
G}{\pi}\left(H^{2}+\frac{k}{a^{2}}\right)\right]
\left(\dot{H}-\frac{k}{a^{2}}\right) = -4 \pi G (\rho+p)
\end{equation}

\begin{equation}
\left(H^{2}+\frac{k}{a^{2}}\right)\left[1~+~\frac{\alpha G}{2
\pi}{\left(H^{2}+\frac{k}{a^{2}}\right)}~\right]=~\frac{8 \pi G
\rho}{3}
\end{equation}

and the continuity equation is

\begin{equation}
\dot{\rho}+3H(\rho+p)=0
\end{equation}

Here $\alpha$ is a dimension less constant. The radius of the
cosmological event horizon is given by

\begin{equation}
{\tilde{r}}_{E}=a\int^{\infty}_{t}\frac{dt}{a}=a\int^{\infty}_{a}\frac{da}{Ha^{2}}
\end{equation}

 The amount of energy crossing the event horizon during time $dt$~is
given by [4,41]

\begin{equation}
-dE=4\pi{{\tilde{r}}_E}^{3}H(\rho+p)dt
\end{equation}

 Now assuming the validity of the first law of thermodynamics at
 the event horizon (i.e. $-dE=T_{E}dS_{E}$) we have

 \begin{equation}
 dS_{E}=\frac{4\pi{{\tilde{r}}_E}^{3}H(\rho+p)dt}{T_{E}}
 \end{equation}

 where $S_{E}$ is the entropy of the event horizon and $T_{E}$ is the temperature on the event horizon.\\

To show the validity of the generalized second law of
thermodynamics we start with Gibbs equation [42]

\begin{equation}
T_{E}dS_{I}=dE_{I}+pdV
\end{equation}

where ~$S_{I}$~ is the entropy of the matter bounded by the event
horizon and ~$E_{I}$~ is the energy of the matter distribution.
Here for thermodynamical equilibrium, the temperature of the
matter inside the event horizon is assumed to be same as on the
event horizon i.e. ~$T_{E}$. Now, starting with

\begin{equation}
V=\frac{4\pi {\tilde{r}_{E}}^{3}}{3} ,~~~~~~E_{I}=\frac{4\pi
{\tilde{r}_{E}}^{3}\rho}{3}
\end{equation}

and the continuity equation (3), the Gibbs equation leads to

\begin{equation}
dS_{I}=-4\pi\frac{{{\tilde{r}}_E}^{2}}{T_E}(\rho+p)dt~,
\end{equation}
where we have used the variation of the radius of the event
horizon [43] as
\begin{equation}
d\tilde{r}_{E}=(\tilde{r}_{E}H-1)dt~
\end{equation}

So combining Eq.(6) and (9) we get

\begin{equation}
\frac{d}{dt}(S_{I}+S_{E})=4\pi
(\rho+p)\frac{{{\tilde{r}}_E}^{2}H}{T_{E}}\left({\tilde{r}}_{E}-\frac{1}H\right)
\end{equation}

Thus the expression in equation (11) for the rate of change of
total entropy show that validity of the second law of
thermodynamics depends both on geometry as well as on the matter
in the universe.

\subsection{\normalsize\bf{\underline{f(R) Gravity}}}

For the spatially flat FRW universe the modified Friedmann
equations [16] in the f(R) gravity are given as

\begin{equation}
8 \pi \rho = \frac{f(R)}{2}-3(\dot{H}+H^{2}-H\frac{d}{dt})f'(R)
\end{equation}

\begin{equation}
8 \pi (\rho+p)=
-2\dot{H}f'(R)+H\frac{d}{dt}f'(R)-\frac{d^{2}}{dt^{2}}f'(R)
\end{equation}

where the Ricci Scalar $R=6\dot{H}+12H^{2}$. The over dot
indicates the derivative with respect to the co-moving time $t$.
The continuity equation is same as equation (3).\\

Since the horizons are determined purely by the geometry,
independent of the gravity theories so as before assuming the
validity of the first law of thermodynamics the change in the
horizon entropy is given by

\begin{equation}
dS_{E}=~\frac{{{\tilde{r}}_{E}}^3H}{2T_{E}}\left(-2\dot{H}f'(R)+H\frac{d}{dt}f'(R)-\frac{d^{2}}{dt^{2}}f'(R)\right)dt
\end{equation}

But this equation can be written also as equation $(6)$. Now from
the Gibbs equation the change in entropy of the matter inside the
event horizon is given by

\begin{equation}
dS_{I}=\frac{1}{T_{E}}(dE_{I}+pdV) =\frac{1}{2T_{E}}
\left[\left(-2\dot{H}f'(R)+H\frac{d}{dt}f'(R)-
\frac{d^{2}}{dt^{2}}f'(R)\right){{\tilde{r}}_{E}}^3
\left({\tilde{r}}_{E}-\frac{1}H\right)dt + \frac{8}3 \pi
{{\tilde{r}}_{E}}^3 d\rho \right]
\end{equation}

Using equation of continuity we get from the expression (13)

\begin{equation}
dS_{I}=~
-\frac{{{\tilde{r}}_{E}}^{2}}{2T_{E}}\left(-2\dot{H}f'(R)+H\frac{d}{dt}f'(R)-
\frac{d^{2}}{dt^{2}}f'(R)\right) dt
\end{equation}

Thus combining Eq.(14) , Eq.(15) and using modified Friedmann
equations we get

$$
\frac{d}{dt}(S_{I}+S_{E})=~\frac{{{\tilde{r}}_{E}}^{2}H}{2T_{E}}\left(-2\dot{H}f'(R)+H\frac{d}{dt}f'(R)-
\frac{d^{2}}{dt^{2}}f'(R)\right)\left({\tilde{r}}_{E}-\frac{1}H\right)
$$

\begin{equation}
= 4\pi(\rho+p) \frac{{{\tilde{r}}_E}^{2}H}{T_{E}}
\left({\tilde{r}}_{E}-\frac{1}H\right)
\end{equation}

 which is same as in the MGLC. So we can conclude that assuming
 the validity of the first law of thermodynamics, the criteria for
 holding the second law of thermodynamics on the event horizon is
 same for the above two modified gravity theories. Further, one
 may note that although for the calculations for $f(R)$ gravity in
 case B we have used the modified Friedmann equations but it is
 possible to do the calculations similar to case A without using
 modified Friedmann equations.\\

\section{\normalsize\bf{Generalized Second Law of Thermodynamics of the Universe Filled with Holographic Dark Energy:}}

The holographic dark energy model is important for the present
accelerating phase of the universe. Also in the present context,
the choice of holographic dark energy is justified as an explicit
expression for the radius of the event horizon can be obtained.Due
to the complicated form of the field equations we consider the
flat FRW model of the universe filled with holographic dark
energy. Now choosing $8 \pi G=1$ we have the modified Friedmann
equation equation in LQG as follows

\begin{equation}
\left(1+\frac{ \alpha}{8
\pi^2}H^2\right)\dot{H}=-\frac{(\rho_{D}+p_{D})}2
\end{equation}

\begin{equation}
H^{2}\left(1+\frac{
\alpha}{16{\pi}^2}H^{2}\right)=\frac{\rho_{D}}3
\end{equation}
with same form of equation of continuity as Eq.(3).\\

The radius of the event horizon considering the holographic dark
energy model [27] is
$$R_{E}=\frac{c}{H\sqrt{\Omega_{D}}}$$
where the density parameter has the expression
$\Omega_{D}=\frac{\rho_{D}}{3H^{2}}$.The equation of state of the
dark energy can be written as

$$
\rho_{D}=\omega_{D}p_{D}$$

where~$\omega_{D}$~is not necessarily a constant.\\
\\

So from the holographic model a small variation of the radius of
the horizon is given by [44]

\begin{equation}
dR_{E}=\frac{3}2R_{E}H(1+\omega_{D})dt
\end{equation}

Since the amount of energy crossing the horizon in time $dt$ does
not depend on any particular gravity theory so from the first law
of thermodynamics we get

\begin{equation}
\frac{dS_{E}}{dt}= 4 \pi H \rho_{D}
\frac{{R_{E}}^{3}}{T_{E}}(1+\omega_{D})
\end{equation}

where $T_{E}$ is the temperature of the event horizon. Now to
obtain the variation of the entropy of the fluid inside the event
horizon we use as before the Gibbs equation (7) with
$$E_{I}=\frac{4}3\pi {R_{E}}^{3}\rho_{t}~~and~~V=\frac{4}3\pi
{R_{E}}^{3}~$$ and using equation (22) we get

\begin{equation}
\frac{dS_{I}}{dt} = \frac{1}{T_{E}}[6 \pi{R_{E}}^{3}H
\rho_{D}{(1+\omega_{D})}^2~ -~ 4 \pi {R_{E}}^{3}H \rho_{D}
(1+\omega_{D})]
\end{equation}

So combining (25) and (26) the time variation of the total
entropy is given by:

\begin{equation}
\frac{d}{dt}(S_{E}+S{I})=6 \pi{R_{E}}^{3}H
\frac{\rho_{D}}{T_{E}}{(1+\omega_{D})}^2
\end{equation}

which is  positive definite i.e. the generalized second law of
thermodynamics is always valid .

One may note that for $f(R)$ gravity if we proceed in the similar
way we have the same conclusion as in Eq.(25). Thus we can say
that for the universe filled with only holographic dark energy the
generalized second law of thermodynamics (GSLT) is always
satisfied on the event horizon assuming the validity of the first
law of thermodynamics. The calculations show that GSLT do not
depend on the modified field equations in these gravity theories
but depend on the
equation of continuity.\\

\section{\normalsize\bf{Generalized Second Law of Thermodynamics of the Universe with Non-Interacting two Fluid System
:}}

In this section we investigate the validity of the generalized
second law of thermodynamics of the universe bounded by the event
horizon when the matter in the universe is taken in the form of
non-interacting two fluid system- one component is the holographic
dark energy model and the other component is in the form of dust.
 The universe is chosen as before to be homogeneous and isotropic and the validity of the
first law has also been assumed here.\\

The modified Friedmann equations in this modified gravity theory
are same as Eq.s (26) and (27) where $\rho$ is replaced by
$\rho_t(=\rho_m+\rho_D)$ and $p$ by $p_D$. Here
$\rho_{D}$~and~$p_{D}$~ correspond to energy density and
thermodynamic pressure of the holographic dark energy model having
equation of state $ p_D= \omega_D \rho_D$ while $\rho_{m}$~is the
energy density corresponding to dust. As the two component matter
system is non interacting so they have separate energy
conservation equations namely

\begin{equation}
\dot{\rho_{m}}+3H(\rho_{m})=0
\end{equation}

and

\begin{equation}
\dot{\rho_{D}}+3H(\rho_{D}+p_{D})=0
\end{equation}

As the universe is bounded by the event horizon so the energy
density of the holographic model can be written as [27]

\begin{equation}
\rho_{D}=3c^{2}{R_{E}}^{-2}
\end{equation}

Taking logarithm on both side and differentiating with respect to
time [assuming $3c^2=1$] we have the small variation of the radius
of the horizon as in Eq.(22).\\

The amount of energy crossing the event horizon in time $dt$~ has
the expression

\begin{equation}
-dE=4\pi{R_{E}}^{3}H(\rho_{t}+p_{D})dt~
\end{equation}

Then the validity of the first law of thermodynamics gives

\begin{equation}
\frac{dS_{E}}{dt}=\frac{4\pi {R_{E}}^3H}{T_{E}}(\rho_{t}+p_{D})
\end{equation}

Now taking
$$E_{I}=\frac{4}3\pi {R_{E}}^{3}\rho_{t}~~and~~V=\frac{4}3\pi
{R_{E}}^{3}~,$$

and using the equation of continuity (37), the Gibbs equation
leads to

\begin{equation}
\frac{dS_{I}}{dt}=\frac{4\pi
{R_{E}}^3}{T_{E}}H(\rho_{t}+p_{D})\left[\frac{3}2(1+\omega_{D})-1\right]
\end{equation}

where we have used the the expression of $dR_{E}$ from Eq.(36).
Hence combining (38) and (39) the resulting change of total
entropy is given by

\begin{equation}
\frac{d}{dt}(S_{I}+S_{E})=\frac{6\pi {R_{E}}^3H}{T_{E}}
(\rho_{t}+p_{D})(\omega_{D}+1)
\end{equation}

Since in the above calculations we have not used any of the
modified field equations, so we have the same result irrespective
of any gravity theory i.e. in MGLC, Gauss-Bonnet gravity etc.

\section{\normalsize\bf{Conclusions:}}

In the present work we examine the validity of the generalized
second law of thermodynamics on the event horizon assuming the
validity of the first law of thermodynamics  in different gravity
theories in a general way. We consider the universe as a
thermodynamical system and is filled with perfect fluid, only
holographic dark energy and non-interacting two fluids in section
II, III and IV respectively. From the above studies we can draw
the following general conclusions in different gravity theories:\\

{\bf I.} If $R_{A}$ and $R_{H}$ denote the radius of the apparent
horizon and the Hubble horizon then the relation between the
horizons are the following[43] :-
$$k=0~(Flat~ model)~:R_{A}=\frac{1}H=R_{H}<R_{E}$$
$$k=-1~(Open~ model)~:R_{H}<R_{A}<R_{E}$$
$$k=-1~(Closed~ model)~:R_{A}<R_{E}<R_{H}$$
$$~~~~~~~~~~~~~~~~~~or$$
$$~~~~~~~~~~~~~~~~~~~~~~~R_{A}<R_{H}<R_{E}$$\\

In section $(II)$ from equation $(11)$ (or $(18)$) we can conclude
that the generalized second law of thermodynamics (GSLT) holds at
the event horizon assuming the first law of thermodynamics
provided the weak energy condition is satisfied for flat and open
model of the universe. However, for closed FRW model, GSLT is
satisfied and $R_H<R_E$ or there is a violation of weak energy
condition (exotic matter) and $R_E<R_H$.\\

{\bf II.} It is to be noted that in reference [43] similar
conclusion was drawn for Einstein Gravity as well as for
Einstein-Gauss-Bonnet (EGB) gravity.\\

{\bf III.}When holographic dark energy is taken as the matter in
the flat FRW universe in section III we have seen that the
generalized second law of thermodynamics holds on the event
horizon assuming the validity of the first law for MGLC and f(R)
gravity. Here we need no restriction either on the
matter or on the geometry.\\

{\bf IV.} Also from section III one can check that if we use the
Gauss-Bonnet theory as a particular case of f(R) gravity then the
Einstein field equation as well as the equations for the
thermodynamical studies are very similar to those that we have
presented for MGLC provided the Gauss-Bonnet coupling parameter
($\tilde{\alpha}$) is related to the dimension less parameter
$\alpha$ in LQG by the relation $\tilde{\alpha}=\frac{\alpha
H^2}{8 \pi^2}$.\\

{\bf V.} In section IV for non-interacting two fluid system, if
the holographic dark energy component individually satisfies the
weak energy condition i.e. the HDE component is not of the phantom
nature then universe as a thermodynamical system with matter in
the form of non-interacting two fluid system always obey the GSLT
in the two gravity theory we are considering. One may note that
identical conclusion was obtained in ref [44]  for Einstein
gravity (see also[45] for EGB gravity).\\

{\bf VI.} One thing may be noted that throughout the paper we have
not used any explicit form of temperature and entropy on the event
horizon.Also we are considering the universe is in thermodynamical
equilibrium so the temperature on the event horizon is similar
with the temperature inside the event horizon.\\

{\bf VII.} To examine the validity of the GSLT on the event
horizon, we have not used any modified Friedmann equations for the
gravity theories (those we are considering) - only the continuity
equation is needed. As Einstein field equations and the first law
of thermodynamics on the apparent horizon are equivalent so we may
speculate that the validity of the thermodynamical laws on the
event horizon does not depend on the validity of the laws at the
apparent horizon.\\

{\bf VIII.} Finally, we may conclude that if we assume the first
law of thermodynamics on the event horizon then validity of GSLT
does not depend on any specific gravity theory (discussed here).
Also the conclusion is not affected by the matter that we have
considered $-$ the only restriction is that any component of the
matter should not be of phantom nature.\\

For further work, we study the validity of GSLT on the event
horizon for other gravity theories and examine whether any general
conclusion independent of any specific gravity theory can be
drawn. Also thermodynamical laws in phantom era may have distinct
features.\\

{\bf References:}\\

$[1]$ T.Jacobson, \it {Phys. Rev Lett.} {\bf 75} 1260 (1995). \\\\
$[2]$ J.D. Bekenstein , {\it Phys. Rev. D} {\bf 7} 2333 (1973).\\\\
$[3]$ T. Padmanavan , {\it Class. Quant. Grav. } {\bf 19} 5387 (2002).\\\\
$[4]$ R.G. Cai and S.P. Kim , {\it JHEP} {\bf 02} 050 (2005).\\\\
$[5]$ S.Perimutter et. al. [Supernova Cosmology Project
Collaboration] \it{Astrophys. J} {\bf 517} (1999) 565; A.G. Riess
et. al. [Supernova Search Team Collaboration] \it{Astron. J} {\bf
116} (1998) 1009; P. Astier et. al. [The SNLS Collaboration]
\it{Astron. Astrophys.} {\bf 447} (2006) 31; A.G. Riess et. al.
\it{Astrophys. J} {\bf 659} (2007) 98.\\\\
$[6]$ D.N. Spergel et. al. [WMAP Collaboration] \it{Astrophys. J
Suppl.} {\bf 148} (2003) 175; H.V. Peiris et. al. [WMAP
Collaboration] \it{Astrophys. J. Suppl.} {\bf 148} (2003) 213;
D.N. Spergel et. al. [WMAP Collaboration] \it{Astrophys. J Suppl.}
{\bf 170} (2007) 377; E. Komatsu et. al. [WMAP Collaboration]
arXiv:0803.0547 [astro-ph].\\\\
$[7]$ V. Sahi , {\it AIP Conf. Proc.} {\bf 782} (2005) 166 ; [J. Phys. Conf. Ser. {\bf 31} (2006) 115].\\\\
$[8]$ T. Padmanavan , {\it Phys. Rept. } {\bf 380} (2002) 235.\\\\
$[9]$ E.J. Copeland, M. Sami and S. Tsujikawa , {\it IJMPD } {\bf 15} (2006) 1753.\\\\
$[10]$ R. Durrer and R. Marteens , {\it Gen. Rel. Grav.} {\bf 40}
(2008) 301.\\\\
$[11]$ S. Nojiri and S.D. Odintsov , {\it Int. J. Geom. Meth. Mod.
Phys.} {\bf 4} (2007) 115.\\\\
$[12]$ S. Nojiri , S. Odintsov , {\it arXiv: 0801.4843} {\bf
[astro-ph]};  S. Nojiri , S. Odintsov , {\it arXiv: 0807.0685}
{\bf [hep-th ]}; S. Capozziello, {\it IJMPD} {\bf 11} (2002) 483.\\\\
$[13]$ S.M.Carroll, V. Duvvuri, M. Trodden and M.S. Turner {\it
Phys. Rev. D} {\bf 68} (2004) 043528.\\\\
$[14]$ S. Nojiri and S.D. Odintsov , {\it Phys. Rev. D} {\bf 68}
(2003) 123512; S. Nojiri and S.D. Odintsov , {\it Phys. Rev. D} {\bf 74} (2006) 086005.\\\\
$[15]$ Kazuharu Bamba and Chao-Qiang Geng , {\it arXiv: 0901.1509}
{\bf [hep-th]}.\\\\
$[16]$ H.M. Sadjadi , {\it Phys. Rev. D} {\bf 76} (2007)
104024.\\\\
$[17]$ C.Rovelli , {\it Phys. Rev. Lett.} {\bf 77} (1996)
3288.\\\\
$[18]$ A. Ashtekar , J. Baez, A. Corichi and K. Krasnov, {\it
Phys. Rev. Lett.} {\bf 80} (1998) 904.\\\\
$[19]$ R.K. Paul and P. Majumder , {\it Phys. Rev. Lett.} {\bf 84}
(2000) 5255.\\\\
$[20]$ A. Ghosh and P. Mitra  , {\it Phys. Rev. D} {\bf 71} (2005)
027502.\\\\
$[21]$ M.Domagala and J.Lewandowski , {\it Class. Quant. Grav.}
{\bf 21} (2004) 5233.\\\\
$[22]$ K.A. Meissner , {\it Class. Quant. Grav.}
{\bf 21} (2004) 5245.\\\\
$[23]$ S.Hod , {\it Class. Quant. Grav.}
{\bf 21} (2004) L97.\\\\
$[24]$ A.J.M. Medved, {\it Class. Quant. Grav.}
{\bf 22} (2005) 133.\\\\
$[25]$ A. Chatterjee and P. Mazumder , {\it arXiv: 0303030}
{\bf [gr-qc]}.\\\\
$[26]$ A. Ashtekar , T. Pawlowski and P. Sing , {\it Phys. Rev.
Lett.} {\bf 96} (2006) 141301; A. Ashtekar ,T. Pawlowski and P.
Sing , {\it Phys. Rev. D} {\bf 74} (2006) 084003; A. Ashtekar ,T.
Pawlowski , P. Sing and K. Vandersloot , {\it Phys. Rev. D} {\bf 75} (2007) 024035.\\\\
$[27]$ B. Wang,Y. Gong,E. Abdalla, \it{Phys. Rev. D} {\bf 74}
 (2006) 083520.\\\\
$[28]$ M. Li , {\it Phys. Lett. B} {\bf 603}  (2004) 01.\\\\
$[29]$ M. R. Setare  and S. Shafei  , {\it JCAP } {\bf 0609}
(2006)  011; arXiv: gr-qc/0606103 .\\\\
$[30]$ B. Hu and Y. Ling \it{Phys. Rev. D} {\bf 73}
(2006) 123510.\\\\
$[31]$ M. Ito , {\it Europhys. Lett. } {\bf 71}  (2005) 712.\\\\
$[32]$ S. Nojiri , S. Odintsov , {\it Gen. Rel. Grav.} {\bf 38}
(2006) 1285 .\\\\
$[33]$ E.N. Saridakis , {\it Phys. Lett. B} {\bf 660}
(2008) 138.\\\\
$[34]$ Q.G. Huang and M. Li , {\it JCAP} {\bf 0408} (2004) 013.\\\\
$[35]$ X. Zhang , {\it IJMPD } {\bf 14} (2005) 1597.\\\\
$[36]$ D. Pavon and W. Zimdahl , {\it Phys. Lett. B} {\bf 628}
(2005) 206.\\\\
$[37]$ H. kim , H.W. Lee and Y.S. Myung , {\it Phys. Lett. B} {\bf
632} (2006) 605.\\\\
$[38]$ S.D. Hsu , {\it Phys. Lett. B} {\bf 594} (2004) 01.\\\\
$[39]$ R. Horvat , {\it Phys. Rev. D} {\bf 70} (2004) 087301 .\\\\
$[40]$ R.G. Cai, L.M. Cao, Y.P. Hu ,arXiv: hep-th/0807.1232 .\\\\
$[41]$ R.S. Bousso, \it{Phys. Rev. D} {\bf 71} 064024 (2005).\\\\
$[42]$ G. Izquierdo and D. Pavon, {\it Phys. Lett. B} {\bf 633}
420 (2006).\\\\
$[43]$ N. Mazumder and S. Chakraborty, {\it Class. Quant. Gravity}
{\bf 26} 195016 (2009).\\\\
$[44]$ N. Mazumder and S. Chakraborty , {\it Gen.Rel.Grav.} {\bf 42} 813
(2010).\\\\
$[45]$ In our earlier work [44] we have studied the generalized
second law of thermodynamics as in section (IV) for Gauss-Bonnet
gravity but we can not make any definite conclusion due to
complicated expression for the rate of change of total entropy. In
the present work we can simplify that expression further and
we have the same conclusion as conclusion (V).\\\\

\end{document}